\documentclass[a4paper,twocolumn,11pt,accepted=2026-04-08]{quantumarticle}
\pdfoutput=1
\usepackage[utf8]{inputenc}
\usepackage[english]{babel}
\usepackage[T1]{fontenc}
\usepackage{amsmath}
\usepackage{amssymb}
\usepackage{hyperref}
\usepackage{graphicx,epstopdf}
\usepackage{svg}

\usepackage{tikz}
\usepackage{lipsum}

\begin{document}

\title{Diversity Methods for Improving Convergence and Accuracy of  Quantum Error Correction Decoders Through Hardware Emulation}

\author{Francisco Garcia-Herrero}
\email{francg18@ucm.es}
\affiliation{Department of Computer Architecture and Automatics, Complutense University of Madrid, Madrid, Spain}
\orcid{0000-0001-6719-9681}
\author{Javier Valls}
\email{jvalls@upv.es}
%\homepage{http://quantum-journal.org}
\orcid{0000-0002-9390-5022}
%\thanks{You can use the \texttt{\textbackslash{}email}, \texttt{\textbackslash{}homepage}, and \texttt{\textbackslash{}thanks} commands to add additional information for the preceding \texttt{\textbackslash{}author}. If applicable, this can also be used to indicate that a work has previously been published in conference proceedings.}
\affiliation{Instituto de Telecomunicaciones y Aplicaciones Multimedia, Universitat Politecnica de Valencia, Valencia, Spain}
\author{Llanos Vergara-Picazo}
\affiliation{Department of Computer Architecture and Automatics, Complutense University of Madrid, Madrid, Spain}
\orcid{0009-0003-8825-5218}
\author{Vicente Torres}
\affiliation{Instituto de Telecomunicaciones y Aplicaciones Multimedia, Universitat Politecnica de Valencia, Valencia, Spain}
\orcid{0000-0002-6829-7889}

\maketitle
\maketitle
\begin{abstract}

\textcolor{black}{As quantum computing moves toward fault-tolerant architectures, quantum error correction (QEC) decoder performance is increasingly critical for scalability. Understanding the impact of transitioning from floating-point software to finite-precision hardware is essential, as hardware decoder performance affects code distance, qubit requirements, and connectivity between quantum and classical control units.}

This paper introduces a hardware emulator to evaluate QEC decoders using real hardware instead of software models. The emulator can explore $\mathbf{10^{13}}$ different error patterns in 20 days with a single FPGA device running at 150 MHz, guaranteeing the decoder's performance at logical rates of $\mathbf{10^{-12}}$, the requirement for most quantum algorithms. In contrast, an optimized C++ software on an Intel Core i9 with 128 GB RAM would take over a year to achieve similar results. The emulator also enables the storage of uncorrectable error patterns that generate logical errors, allowing for offline analysis and the design of new decoders.

Using results from the emulator, we propose a method that combines several belief propagation (BP) decoders with different quantization levels, which we define as a diversity-based decoder. Individually, these decoders may show subpar error correction, but together they outperform the floating-point version of BP for quantum low-density parity-check (QLDPC) codes like hypergraph or lifted product. Preliminary results with circuit-level noise and bivariate bicycle codes suggest that hardware insights can also improve software. Our diversity-based proposal achieves a similar logical error rate as the well-known approach, BP with ordered statistics (BP+OSD) decoding, with average speed improvements ranging from 30\% to 80\%, and 10\% to 120\% in worst-case scenarios, while reducing post-processing algorithm activation from 47\% to 96.93\%, maintaining the same accuracy.

\end{abstract}

% Make the title area
\maketitle

%%%%%%%%%%%%%%%%%%%%%%%%%%%%%%%%%%%%%%%%%%%%%%%%%%%%%%%%%%%%%%%%%%%%%%%%%%%%%%%%
\section{Introduction}

The design of decoding algorithms that obtain high accuracy and low latency for quantum low-density parity-check (QLDPC) codes \cite{babar2015fifteen} has been a very active area during the last decade, especially over the previous six years.

Researchers have been working on low-complexity decoding algorithms to achieve the performance of belief-propagation (BP) combined with ordered-statistics decoding (OSD) \cite{roffe2020decoding}, \cite{panteleev2021degenerate}. This widely used approach offers good accuracy, but can be too slow for real-time decoding.

BP decoders are attractive due to their low complexity and high parallelization for real-time hardware implementations reaching low latency and allowing scalable architectures \cite{valls2021syndrome}, \cite{müller2025improvedbeliefpropagationsufficient}. However, BP introduces degradations on accuracy that are primarily due to the properties of QLDPC codes, such as degeneracy \cite{Fuentes2021Degeneracy}. Degeneracy refers to the existence of multiple error patterns that have the same effect on the logical information, which standard BP decoding fails to properly account for. In addition, the abundance of low-weight stabilizers in QLDPC codes contributes to the formation of trapping sets, which are specific substructures in the code’s Tanner graph that do not allow messages to propagate properly through the graph. These substructures are more prevalent in QLDPC than in classical LDPC codes \cite{raveendran2021trapping}. Trapping sets can cause the decoder to converge to incorrect solutions or oscillate between inconsistent ones, limiting the overall decoding performance. The trapping sets are one of the major contributors to the emergence of an error floor, a phenomenon where the decoder's logical error rate (LER) stops decreasing exponentially with the improvement of the physical error rate, thus limiting accuracy. 

To reduce the degradation in the LER produced by BP, OSD is used as a post-processing method, improving accuracy by several orders of magnitude \cite{roffe2020decoding}, \cite{panteleev2021degenerate}. However, the primary drawback of OSD is that it must solve a dynamically created linear system based on the reliability information provided by BP to obtain a solution. This process of solving the linear system can create bottlenecks for implementing parallel architectures, since it involves sequential operations and dynamic computation of matrix inverses. These factors lead to scalability challenges, particularly as the size of the linear system to be solved increases, especially when considering the detector error model \cite{derks2024designing}, \cite{Gidney2021stimfaststabilizer}.

Some alternative decoders focus on avoiding OSD-like solutions that require solving systems of linear equations, which implies the computation of matrix inversion through Gaussian elimination. These alternatives aim to improve efficiency by combining different BP decoders or modifying the internal update rules of the variable and check nodes, based on the expectation that each variation enables the correction of distinct error patterns that are not successfully handled by a single BP decoder. Examples of these include Relay-BP \cite{müller2025improvedbeliefpropagationsufficient}, Belief Propagation Guided Decimation (BPGD) \cite{yao2024belief}, \cite{gong2024lowlatencyiterativedecodingqldpc}, Stabilizer Inactivation (SI) \cite{du2022stabilizer}, and Check-Agnosia (CA) \cite{du2024check}. 

On the other side, some decoding algorithms aim to reduce the complexity of the post-processing method, rather than using only BP decoders. Some examples are the Localized Statistics Decoder (BP+LSD) \cite{hillmann2024localizedstatisticsdecodingparallel}, the Ordered Tanner Forest (BP+OTF) \cite{iolius2024almostlineartimedecodingalgorithm}, and  Ambiguity Clustering (BP+AC) \cite{wolanski2025ambiguityclusteringaccurateefficient}, which obtain high accuracy while maintaining matrix inversion to perform the decoding. 

Regardless of the approach taken, all solutions begin with some variant of BP, leading to further exploration of BP algorithms to improve accuracy and reduce complexity and hence latency. This includes modifications to BP's scheduling, such as the random reordering of node updates \cite{du2023layered}, or introducing additional noise or perturbations to address the error floor problem \cite{Poulin2008Onthe}.

However, several important questions remain unanswered regarding the QEC decoding algorithms. First, most quantum algorithms require a LER between $10^{-10}$ and  $10^{-13}$ \cite{kim2024faultolerant}. To ensure that, it is necessary to know the impact of finite-precision architectures on the final error correction performance of the decoders when implemented in hardware, especially in the low LER regions. To obtain statistically significant results, at least $10^{12}$ or $10^{15}$ experiments need to be run, which would be a bottleneck in software, which would require thousands of CPUs to finish the simulations within months. On the other hand, given the scaling challenges that the classical hardware will need to face to implement the QEC layer of a fault-tolerant quantum computing, it is reasonable to believe that floating-point operations may not be power or time-efficient, and accuracy should be evaluated with this limitation in mind \cite{kadomoto2024preliminary}. 

Answering these questions requires the implementation of an emulator. The scope of the emulator is to evaluate and benchmark QEC decoders in real hardware at low LERs, enabling large-scale error pattern sampling, accuracy validation, and decoder optimizations beyond the limits of software simulation. Although designing and rigorously verifying such an emulator is time-consuming, it is essential to address with warranties some of the previous problems. The accuracy of the final device will ultimately depend on the hardware architecture rather than the software implementation. The verification of the accuracy of the implemented decoders in the region of lower LERs cannot be realistically achieved through alternative methods, given limited resources. Additionally, the effects of quantization noise in finite precision implementations may lead to the development of new decoders that take advantage of the perturbations introduced by these architectures.

Finally, it is important to evaluate real hardware implementations of BP, which serves as the core algorithm behind all the aforementioned state-of-the-art decoders \cite{roffe2020decoding}, \cite{panteleev2021degenerate}, \cite{müller2025improvedbeliefpropagationsufficient}, \cite{yao2024belief}, \cite{gong2024lowlatencyiterativedecodingqldpc}, \cite{du2022stabilizer}, \cite{du2024check}, \cite{hillmann2024localizedstatisticsdecodingparallel}, \cite{iolius2024almostlineartimedecodingalgorithm}, \cite{wolanski2025ambiguityclusteringaccurateefficient}. Even if other approaches, such as those based on neural networks and BP, are explored \cite{Gong2024Graph}, utilizing a hardware emulator will enable more accurate training for achieving LERs below \(10^{-12}\) (as the samples will be finite precision too), which is currently not feasible with software-based methods.

%%%%

\textcolor{black}{The main contributions of this paper are
summarized as follows:}

\begin{itemize}

\item {The architecture and implementation of a hardware emulator to benchmark QEC real-time decoders and to facilitate offline analysis of the derived information. This emulator enables a significant reduction in the time required to explore low error rates down to $10^{-12}$, allowing this process to be completed in days instead of the years it would take using a software model on a CPU.}\\

\item{The design of a BP-based decoder with the knowledge obtained after studying, through the emulator, the diverse levels of noise generated by the quantization schemes. This proposal improves the LER without adding sequential post-processing algorithms like OSD. Similarly to the one in \cite{shutty2024efficientnearoptimaldecodingsurface}, in which the authors combined several noisy decoders to generate highly accurate decoding predictions. However, in our proposal, the noisy decoder is BP instead of minimum-weight perfect matching (MWPM), and the noise is not added artificially but is part of the nature of the hardware architectures, which are carefully selected. %Finally, due to the nature of BP, it is not necessary to have a degree of consensus among the decoders to increase confidence; we only wait for the one that converges first, following some priorities in between the different quantization schemes.
Our solution is based on just four decoders that can share the hardware in groups of two without impact on latency and a minimum memory overhead. The codes under test are hypergraph product and lifted product QLDPC codes \cite{panteleev2021degenerate}, \cite{raveendran2022soft}. These codes were selected due to their extensive documentation in the literature under the code capacity model, which allows for verification of the emulator's accuracy.}\\

\item{The definition of a decoder that successfully reduces both the number of BP iterations and the frequency of post-processing calls under circuit-level noise. At the same time, it maintains or improves the LER when compared to BP+OSD on bivariate bicycle codes of varying lengths and distances \cite{bravyi2024high}.}\\

\end{itemize}

\textcolor{black}{The structure of the paper is detailed as follows: Section II describes the proposed emulator, detailing the components of the architecture and the different configuration parameters; and introduces the verification of the platform by testing the performance of several QLDPC codes with check-node degrees of six and eight, from which some interesting conclusions are drawn. Section III presents the first diversity approach that improves the accuracy of BP decoders with moderate overhead in hardware resources. This approach leverages the quantization noise introduced by finite-precision architectures to achieve low-latency solutions. Section IV introduces a second diversity proposal based on different BP implementations, which can lead to reduced hardware requirements and shorter execution times by minimizing the number of post-processing executions. Finally, Section V concludes the paper and suggests future research lines.}

\section{Proposed Emulator}

In this section, we outline the general architecture of the emulator and its schedule, along with the main parameters measured for post-analysis of the simulations. This analysis will provide a better understanding of how decoders behave with finite precision. The insights gained will help to: i) customize the design of the decoders; and ii) develop more efficient post-processors that improve decoding accuracy while reducing overhead in area, power consumption, and latency.

\begin{figure*}[!t]
\centering
\includegraphics[angle=-90, width=\dimexpr1.1\textwidth\relax]{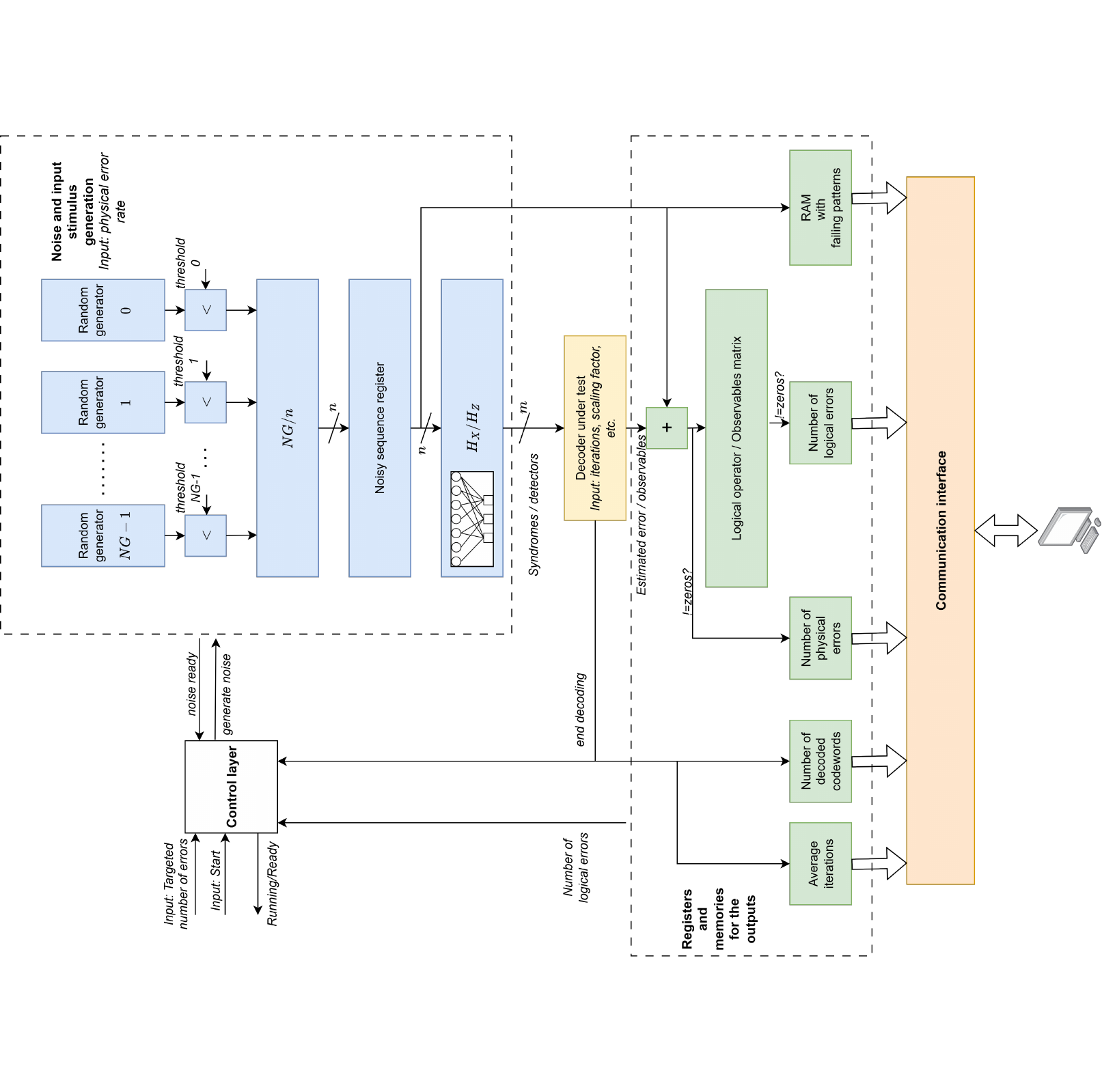}
\caption{Simplified diagram of the proposed emulator's architecture.}
\label{fig:GeneralEmulator}
\end{figure*}

The architecture depicted in Fig. \ref{fig:GeneralEmulator} consists of three main layers: noise and input stimulus generation (highlighted in blue in the figure), an input/output parameter interface (green), and a communication interface (orange). Additionally, there is a transversal control layer that orchestrates the entire emulator. This structure is adaptable for any type of QEC decoder (highlighted in yellow in Fig.\ref{fig:GeneralEmulator}) that utilizes syndromes or detectors \footnote{A detector is a parity constraint on a set of measurement outcomes, as define in \cite{derks2024designing}.} as inputs and produces estimated error vectors or observables as outputs. As an example, Section 2.6 analyzes the particular case of evaluating BP decoders with the emulator.

It is important to note that the implementation is vendor-agnostic, meaning it can be executed on any FPGA device, and it does not rely on proprietary IP cores \footnote{An Intellectual Property (IP) core is a reusable, modular block of hardware that implements a specific function that can be integrated into a larger hardware design to accelerate development and ensure an optimize performance when implemented on an specific technology.}.

Next, we detail each of the modules. \textcolor{black}{The code of the corresponding hardware description language of the main modules can be found in \href{https://github.com/fpgaqec/QLDPC_FPGA_Emulator/}{https://github.com/fpgaqec} with an example for a short QLDPC code (with a 127x254 parity-check matrix). The code and architecture are fully generic, requiring only minor modifications to parameters and constants. Please note that constraint files, which map the code to hardware, are device-dependent.}

\subsection{Noise and input stimulus generation}
The architecture consists of $NG$ noise generators. Each noise generator is equipped with a Gaussian noise generator \cite{Guiterrez2012Hardware} that produces a sequence of values between 0 and 1, with an accuracy of 18 bits and a configurable seed. This design allows simulations to be replicated if it is necessary to reproduce specific patterns. Additionally, using a configurable seed helps prevent overlapping sequences, making it easier to distribute experiments across multiple boards.

At the output of each of these generators, a comparator is implemented to determine whether an X, Y, or Z error has occurred at a given location. This comparator has one of the inputs wired to a programmable threshold, whose value depends on the error model and the physical error rate. 

Depending on the FPGA device and the length of the error vector $n$, which is equal to the codeword length for the code-capacity model, having $NG=n$ random noise generators may consume too many hardware resources, leaving insufficient space for the QEC decoder. The parameter $NG$ is generic and configurable at synthesis time. After the comparators, there is one register that receives $NG$ bits in parallel and outputs $n$ bits. So, when $NG<n$, the number of clock cycles to generate an entire error pattern is $\left \lceil{n/NG}\right \rceil$. When this error sequence is generated, it is stored in a parallel register to start producing the next noise sequence while the decoder is working. This minimizes idle periods.

After the register is updated with a new sequence, the syndrome or detector pattern is computed by multiplying by the corresponding parity check matrix ($H_X$ or $H_Z$) or the graph associated with the detector error model. This binary product is performed in parallel and transmitted to the decoder under test. This is the last step of the module and has a total latency of $\left \lceil{n/NG}\right \rceil+3$ clock cycles.

\textcolor{black}{Supporting both $H_X$ and $H_Z$ simulations simultaneously for realistic codes, such as the (144,12,12) QLDPC code, would require a multi-FPGA setup \cite{müller2025improvedbeliefpropagationsufficient}. Even the largest FPGA devices available in 2025 (Virtex UltraScale+ V19P) can accommodate a decoder for only one type of error, with the layout nearly full, and adding real-time noise generation and the rest of the emulator parts would further increase resource demands and require precise synchronization. While the ideal setup would handle both error types concurrently, this does not affect the main findings of this work on the impact of quantization on decoding accuracy. Multi-FPGA implementations also involve substantial complexity, including multiple high-performance evaluation boards and synchronization protocols (e.g., Aurora or PCIe). We therefore note this as future work, as this study represents a first step toward a complete setup that could also consider decoders for correlated errors.}

\subsection{Decoder} The emulator is compatible with any QEC decoder that accepts syndromes or detectors as inputs and generates errors as outputs. The interface with the rest of the emulator is straightforward; only start, ready, and done signals are needed to control the decoder. Additional information, such as the number of iterations of the decoder, can be computed in the control layer or directly provided by the decoder. 

As an example, we have evaluated the fully parallel architecture for the scaled min-sum algorithm described in \cite{valls2021syndrome}, applying it to multiple codes from various code families, which we will detail in the following sections.  As a summary, the scaled min-sum algorithm is a BP-based message-passing algorithm that operates on a Tanner graph with variable and check nodes. Variable nodes represent potential error locations and combine prior information from the error model with incoming messages from check nodes, and a scaling factor $\alpha$ that is applied to the check-to-variable messages to increase accuracy. Check-nodes compute their output messages depending on whether the input messages coming from the variable nodes satisfy the syndromes or not. A hard decision (or binary decision) is made at each variable node to determine whether an error is present at that location. Decoding succeeds if this decision satisfies all syndrome constraints.

\subsection{Input/output parameter's interface}
The emulator receives a series of parameters to set up the noise and the accuracy of the simulation, which can be configured at run time, avoiding the need to go through the entire synthesis and place-and-route processes. Additionally, several parameters can be queried during runtime to check how the simulation is evolving or which configuration parameters have been used. This is interesting when low LERs are explored, especially for long codes, when the simulations can reach daytime duration if just one FPGA device is used. The input parameters are:

\begin{itemize}
    \item The physical error rate to be evaluated, with a precision of 18 bits.
    \item Decoder specific parameters. In the case of BP-based decoders, the maximum number of iterations. It was limited to 8 bits, but the number of bits associated can be modified at synthesis time.
    \item Targeted number of errors. To determine statistical significance in a simulation, designers typically set the number of Monte Carlo simulations to $100/\text{LER}$. However, because predicting the exact LER can be challenging due to potential error floors, we employ a different approach. Instead of performing a fixed number of Monte Carlo simulations, we stop the emulation based on a maximum number of uncorrected error patterns, i.e., 100 uncorrected error patterns. The simulation will continue until this maximum is met. 
    
    With 16-bit precision, each simulation can uncover up to $2^{16}$ error patterns. This method allows us to generate datasets that contain a large and diverse array of error patterns. These datasets can later be analyzed offline to enhance decoding algorithms, fine-tune parameters associated with the decoder, or provide sufficient samples for training a neural network-based decoder \cite{Gong2024Graph}.
    \item The start signal. Although it is not a real parameter, it is defined in the interface to be activated externally via software. This signal initiates the whole emulator's control unit.
\end{itemize}

The output parameters are:

\begin{itemize}
    \item The number of physical errors. It is represented by a 16-bit counter that increments when the error vector produced by the decoder does not match the output from the noise generators.
    \item The number of logical errors. It is also a 16-bit counter that is increased if a logical error occurs. To check that, the estimated error sequence computed by the decoder is combined (XOR-ed) with the real error sequence and multiplied by the logical operator matrix to compute the number of logical errors. The same is performed with the output observables and the observable matrix, if the detector error model is implemented. The binary products and the XOR operation are computed in parallel to avoid adding extra latency to the computation.
    \item The number of decoded frames (successfully or not). It is an 80-bit counter that accumulates the number of simulations that are performed for a given configuration. This value is oversized as it can compute a LER of $10^{-23}$.
    \item The total number of iterations of the decoder (if iterative). It is also a counter that increases according to the information provided by the decoder under test about the number of iterations run. It is very useful to estimate the average number of iterations after days of simulations and infer the real-time average latency of the decoder. 
    \item The input parameters. All the previously mentioned parameters — such as the physical error rate, the number of errors to be found, and the maximum number of iterations — can also be accessed to verify the settings configured for the currently running simulation. This can help identify a simulation when multiple boards are operating in parallel.
    \item The running and ready (done) signal. When the simulation has started, the running signal is active to indicate that parameters cannot be modified without a reset. The ready signal indicates that we have found the number of errors that were fixed during the configuration.
\end{itemize}

\begin{figure*}[!t]
\centering
\includegraphics[width=\textwidth]{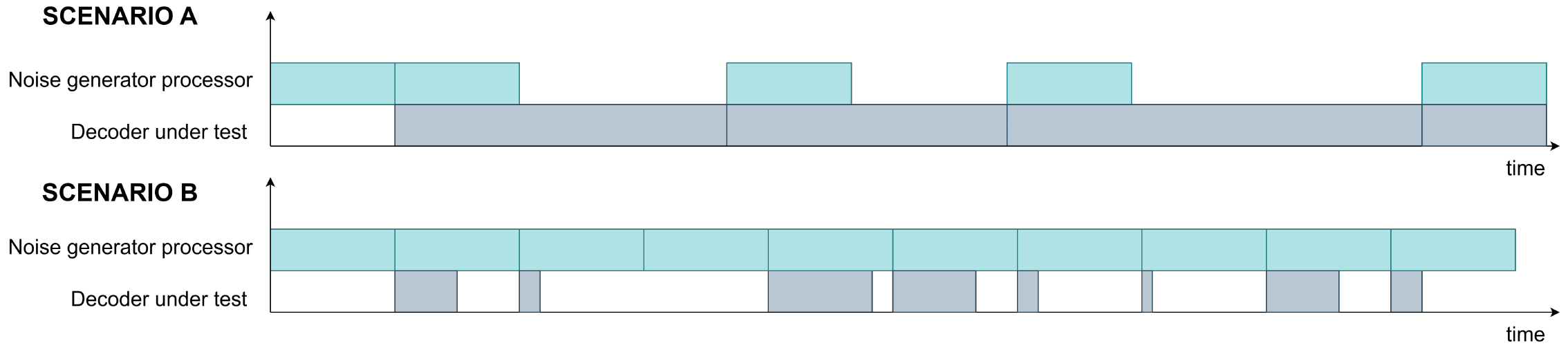}
\caption{Schedule of the proposed emulator. We can see two scenarios: A) when there is a high level of noise and the decoder needs multiple iterations to converge, so it is slower than the noise generator; B) when there is a low level of noise and the decoder converges faster than the next generation of noise. Blue and gray colors indicate when the noise generator processor and the decoder are active, respectively.}
\label{fig:ControlEmulator}
\end{figure*}

\subsection{Communication interface} To reduce dependence on the selected FPGA and minimize the time needed to recover information from simulations, we implemented a full stack based on Gigabit Ethernet. This includes layers for UDP, ARP, MAC, and SGMII interfaces. To enhance flexibility and provide protection against unsolicited or spurious network traffic when the FPGA operates within a shared network environment, we designed a hardware layer implementing a customized protocol to filter and control incoming messages. This protocol supports messages that facilitate communication with the entire emulator during and after experiments.

The primary operations supported by this protocol include starting and stopping the emulator, as well as reading and writing various parameters, such as physical error rate, LER, total number of runs, average number of iterations, and clock cycles. Messages that do not conform to the protocol or are not directed to the FPGA's MAC address are automatically filtered out and rejected at the hardware level (link layer).

One important feature of this architecture is its capability to recover all error patterns that could not be successfully decoded. This allows for offline analysis, enabling the development of more accurate and efficient decoders.

\subsection{Control layer} The control unit responsible for coordinating the emulator's workflow is based on two finite state machines. The first machine orchestrates the generation of noise samples, while the second controls the activation of the decoder under test. Typically, QEC decoders, such as BP decoders, incorporate an early stopping criterion. Therefore, it is essential to notify the noise generators when the decoder has finished to ensure that the next noise sample is ready. In the same way, the decoder must be informed of the noise generator's status to confirm that the noise samples are available.

This communication is particularly crucial when the number of noise generators $NG$ is less than the number of required noise samples $n$. In such cases, $\left \lceil \frac{n}{NG} \right \rceil$ cycles are necessary, and the decoder may finish before the next round of samples is generated due to the early stopping criterion (see Scenario B in Fig. \ref{fig:ControlEmulator}). 

Counterintuitively, in regions where the physical error rate is reduced, the time bottleneck can occur during noise generation. This is because the decoders require a large amount of area resources and leave insufficient space on the FPGA, even for the largest existing devices, to introduce enough noise generators to perform all the computations in parallel. 

We can generally categorize two scenarios: one in which the decoder consumes more clock cycles than the noise generation (Scenario A in Fig. \ref{fig:ControlEmulator}), and another where the decoder is faster (Scenario B). For instance, in a fully parallel BP decoder, the dividing line between both scenarios occurs when the number of iterations required for convergence is fewer than $\left \lceil \frac{2n}{NG} \right \rceil $, assuming two clock cycles per iteration, one for computing the check nodes and another for computing the variable nodes.

The signals exchanged between both finite state machines include ``noise ready'', ``generate noise'', and ``end of decoding''. For simplicity, other control signals, such as those used to store error patterns in RAM when a logical failure occurs (for subsequent a posteriori analysis), have been omitted.

\begin{figure*}[t!]
\centering
\includegraphics[width=0.65\textwidth]{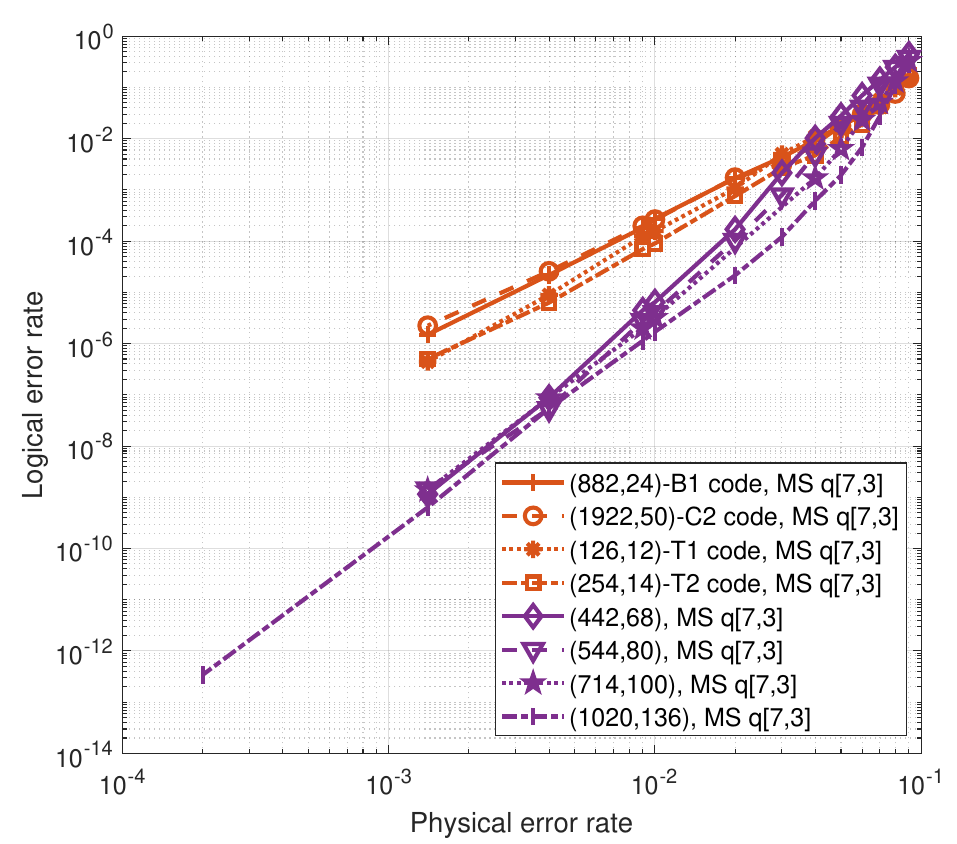}
\caption{LER simulations for eight QLDPC codes of degrees 6 and 8 obtained with the implemented emulator.}
\label{fig:FirstEmulatorResults}
\end{figure*}

\subsection{Results for BP decoders}
The emulator discussed in the previous section was tested with various QLDPC codes, which are defined as ($n$, $k$, $d$), where $n$ is the number of physical qubits, $k$ is the number of logical qubits, and $d$ is the minimum distance of the code. Fig. \ref{fig:FirstEmulatorResults} presents a small sample of eight codes with check node degrees\footnote{The check node degree is the number of messages computed by a check node in a QLDPC code.} of six and eight, decoded using the scaled min-sum (MS) algorithm quantized to 7 bits, with 3 of those bits for the fractional part. The codes examined include hypergraph product codes and lifted product QLDPC codes, as studied in \cite{panteleev2021degenerate} and \cite{raveendran2022soft}, respectively \footnote{The code names B1, C2, T1, and T2 are maintained from references \cite{panteleev2021degenerate,raveendran2022soft} to help readers identify the corresponding construction methods in the original works.}. Specifically, the hypergraph product codes with check node degree six are B1 (882, 24, $\leq$ 24) and C2 (1922, 50, 16). The lifted product codes with check node degree eight include (442, 68, $\leq$ 10), (544, 80, $\leq$ 12), (714, 100, $\leq$ 16), and (1020, 136, $\leq$ 20). Additionally, two codes from \cite{du2022stabilizer}, T1 (126, 12, < 11) and T2 (254, 14, $<$ 17), are also analyzed.

The emulator was implemented on an AMD Virtex UltraScale+ FPGA VCU118 \cite{amd_vcu118}, but it is worth noting that it can be ported to any FPGA with sufficient logic resources. The evaluated architecture for BP utilized a fully parallel MS algorithm with a flooded scheduling approach from \cite{valls2021syndrome}. The maximum frequency for the decoders was set at 150 MHz, with two clock cycles required per iteration, equating to 13.3 ns per iteration. The number of noise generators was consistently set to $NG=40$ across all experiments, resulting in a noise generation latency ranging from 4 clock cycles for the shortest code (T1, $\left \lceil \frac{126}{40} \right \rceil =4$) to 49 clock cycles for the largest one (C2, $\left \lceil \frac{1922}{40} \right \rceil =49$). In these experiments, the thresholds of the noise generators were configured to emulate a code-capacity model.

\begin{figure*}[t!]
\centering
\includegraphics[width=0.65\textwidth]{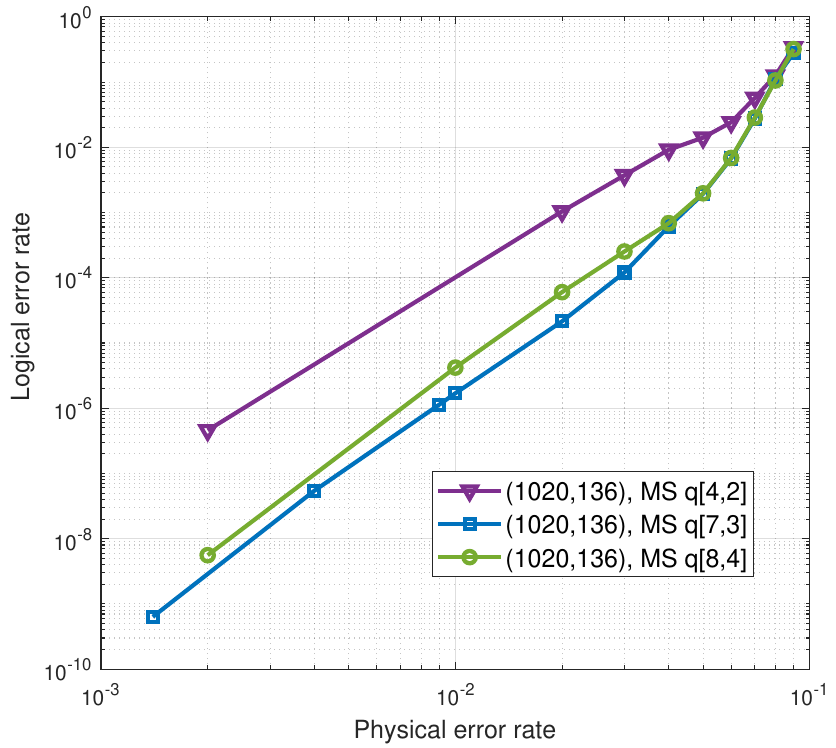}
\caption{Effect of various quantization schemes on the LER for a QLDPC code with a check node degree of 8 obtained with the proposed emulator.}
\label{fig:QuantEmulatorResults}
\end{figure*}

For the shortest code, the latency of noise generation would equate to approximately 2 iterations of the MS decoder, while in the worst case, it could reach 13 iterations due to the latency introduced by the noise generators. Running all simulations at a physical error rate ranging from $10^{-3}$ to $10^{-1}$ on VCU118 took 4 hours and 48 minutes in the worst-case scenario, achieving a LER of $10^{-9}$. For the (1020, 136) QLDPC code, emulating more than $10^{13}$ samples required 20 days. Faster simulations could be achieved by increasing $NG$ or using multiple boards. To replicate the same number of scenarios using the efficient software library from \cite{Roffe_LDPC_Python_tools_2022} on an Intel Core i9-14900KF \cite{intel_i9_14900kf} with 128GB of RAM, the estimated time would exceed one year. Although this result could potentially be improved by increasing parallelism or utilizing a high-performance computing (HPC) system, even with that, the simulation would not be able to replicate the exact behavior of the hardware, which is the aspect that we will also try to exploit, as we will explain in the next sections.

Beyond speed, our experiments yielded other important insights. For example, the impact of distance on the codes with degree eight (represented in purple) diminishes after a LER of \(10^{-5}\), where there is no aggressive error floor. Additionally, we observe distinct trends between QLDPC codes with different degree distributions. The codes with degree six exhibit a significant degradation with BP, necessitating a physical error rate that is extremely low to achieve a LER below \(10^{-12}\), making them challenging to decode without post-processing. Conversely, the degree eight codes reach the target LER before encountering a physical error rate of \(10^{-4}\). It is obvious that these codes face different limitations, such as the connectivity of degree eight on the quantum side \cite{bravyi2024high}, but the behavior decoded looks interesting, and as we will discuss in the next section, their error correction performance can be improved just by using BP.

In addition to the previous results, we analyzed the effect of quantization noise not only to understand its impact on LER performance and hardware savings in terms of area, power, and time but also because some studies have shown that a certain level of noise can enhance the convergence of BP decoders \cite{Poulin2008Onthe}, \cite{du2023layered}, \cite{du2024check}. For this reason, we tested different quantization schemes for the messages computed in BP to determine if we could benefit from the quantization noise.

Some of the quantization schemes we tested include 8 bits, with 4 of them to represent the fractional part of the messages (q[8,4]); 7 bits, with 3 of them fractional (q[7,3]); and an extreme case of 4 bits, with 2 fractional (q[4,2]). Fig. \ref{fig:QuantEmulatorResults} presents a comparison of the LER obtained with a BP decoder for the (1020,136) QLDPC code applying different quantization schemes. For the sake of simplicity, we omitted the results for the other codes and quantization schemes, but similar conclusions were drawn.

As observed, the most trivial outcome is that using only 4 bits significantly degrades performance compared to using 7 or 8 bits. Additionally, the performance difference until a LER of $10^{-4}$ seems reasonable; thus, 7 or 8 bits appear sufficient for message exchange between check nodes and variable nodes \cite{Hailes2016Survey}. However, a surprising finding is that below a LER of  $10^{-4}$, the scheme with fewer bits outperforms those with more bits, indicating that quantization noise may assist in the convergence of the algorithm. \textcolor{black}{The observation that lower bit-width quantization can outperform higher precision at low logical error rates (below $10^{-4}$) is not new to the decoding literature. In classical LDPC decoding \cite{ngassa2015density}, \cite{mheich2016code}, \cite{zhang2013quantized}, \cite{richardson2003error}, prior works have demonstrated that coarse quantization limits message magnitudes, suppresses harmful feedback loops, and mitigates trapping-set effects, especially in the error-floor regime. However, quantum LDPC codes differ from their classical counterparts due to degeneracy and strong correlations in their graph structure \cite{panteleev2021degenerate}, which invalidate traditional analytic methods such as density evolution. As a result, hardware-level emulation becomes essential for accurately assessing decoding performance under finite-precision constraints. In this context, quantization introduces structured noise that disrupts feedback loops and prevents cycles from reinforcing erroneous messages, which can improve decoding accuracy. While further theoretical development is needed to predict the most effective quantization schemes without exhaustive emulation, this phenomenon highlights the need for innovative approaches to analyze quantum LDPC decoders, where classical methods fall short. In the meantime, emulation remains the only reliable way to verify the behavior of quantum LDPC decoders under finite-precision constraints.}

Exploiting this noise that comes out naturally in hardware implementations could be beneficial and worth exploring, as opposed to other proposals that artificially introduce noise \cite{du2023layered}, \cite{Poulin2008Onthe}, and \cite{shutty2024efficientnearoptimaldecodingsurface}. This raises a critical question: Do different quantization schemes in the decoder lead to the correction of a different set of error patterns due to biases introduced by quantization noise? In other words, are we correcting the same set of error patterns with a larger number of bits as we do with fewer bits, or are we just correcting more errors but of a different set?  This is the starting point for our diversity method proposal, described in the next section.

\section{Diversity based on quantization noise}

One of the main advantages of using the implemented emulator is its ability to store the error patterns that lead to logical errors after decoding with real hardware, especially for low LERs. This capability is significant because, as discussed in the previous section, having more bits does not necessarily mean there will be fewer errors. In other words, certain error patterns that a floating-point decoder cannot correct may be successfully addressed by a finite precision decoder that uses fewer bits. Therefore, capturing the patterns that cause decoding failures in real hardware is essential for effective analysis.

The first step was to fix a specific seed in the random noise generators to ensure the same scenario and perform identical simulations (with the same inputs) for the different quantization schemes. We then captured the patterns of failure and analyzed their differences when decoders with different quantization schemes were applied offline. Our initial conclusion from this experiment was that the sets of failure patterns were different by a significant percentage, confirming our hypothesis that different quantization schemes introduce varying levels of noise that affect decoding in distinct ways. Additionally, we observed that quantization schemes using fewer bits often forced convergence, frequently resulting in incorrect codewords. This observation was fundamental, as it is preferable to experience a larger number of decoding failures than to converge on incorrect codewords because otherwise, errors may go undetected.

Considering the previous observations, we establish a prioritized chain of decoders. The most accurate decoder, q[7,4], will be executed first, followed by q[8,4], which provides a balance between correcting patterns not handled by the previous decoder and detecting decoding failures. Finally, we will use the less accurate decoders, q[4,2] and q[3,1]. The process will stop as soon as we achieve convergence, which saves both time and power. 

Additionally, it is important to note that the total number of bits used is less than 32 (the typical width of single-precision floating-point operations on a general-purpose CPU). This reduction not only enhances the speed of implementation compared to software but also improves efficiency in terms of power consumption.

As an example, with the code (1020,136) and a physical error rate close to $10^{-3}$, decoders that function as post-processors (MS with q[8,4], q[4,2], and q[3,1]) are activated only $10^{-7}\%$ of the time. Among these activations, the decoder with the highest accuracy succeeds in $95\%$ of the calls, while the one with the lowest accuracy only succeeds $3.5\%$ of the time. When the physical error rate is closer to $10^{-2}$, the most accurate decoder succeeds in just $78\%$ of the cases, and the second decoder succeeds $16\%$ of the time. In this scenario, the post-processor is called upon $10^{-4}\%$ of the time.

\begin{figure*}[t!]
\centering
\includegraphics[width=0.85\textwidth]{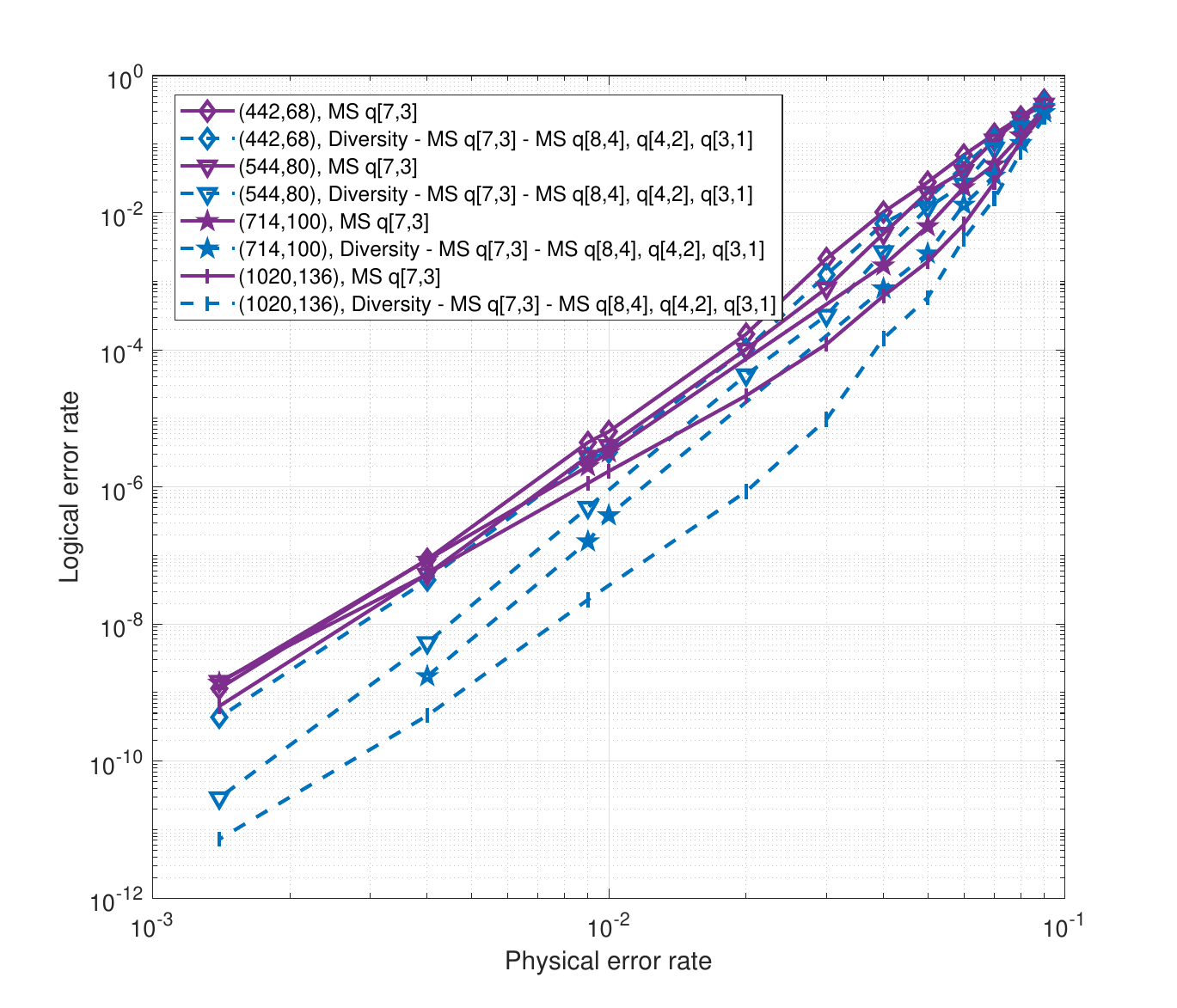}
\caption{LER simulations for four QLDPC codes of degree 8 obtained with the proposed emulator and the diversity approach based on quantization noise.}
\label{fig:DiversityEmulatorResults}
\end{figure*}

This approach was applied to all the lifted product codes of degree eight discussed in the previous section. In Fig.\ref{fig:DiversityEmulatorResults}, we summarize the results, which show improvements for all codes. Additionally, some gain derived from the code distance is partially recovered in the LER region below $10^{-10}$. In some instances, the gain in LER exceeds one order of magnitude when the physical error rate improves.

As we will discuss in the next section, our objective is not to replace other post-processors like LSD or OSD with the diversity decoder if the same accuracy cannot be warranted only with BP-based decoders \cite{crest2024blindnesspropertyminsumdecoding}, but rather to minimize the number of calls, as we will demonstrate later.

The results presented here are just a preliminary approximation; further analysis is needed to optimize these quantization schemes, which will play an essential role when we transition to FPGA or ASIC implementations as quantum systems scale up and more efficient solutions are sought.

\section{Diversity based on BP implementations}
The decoder described in the previous section operates under code-capacity model. In this section, we aim to extend the idea of the diversity decoder to circuit-level noise, verifying it through a software implementation based on Stim \cite{Gidney2021stimfaststabilizer} and the software library from \cite{Roffe_LDPC_Python_tools_2022}. Instead of introducing diversity through quantization noise, we achieve it by varying the scaling factor, modifying the a priori information, and combining BP decoders with different update rules. Our approach includes providing feedback to subsequent decoders if the previous ones do not achieve convergence. Unlike other methods, such as stabilizer inactivation or check-agnosia proposed in \cite{du2022stabilizer} and \cite{du2024check}, we only modify the a priori information based on the binary outputs of the previous decoder (the hard decisions of the estimated error vector), intentionally avoiding sorting steps or graph analysis, which are computationally expensive when the detector error model is applied. When diverse BP has a decoding failure, we apply a post-processor like LSD or OSD, but we significantly reduce the number of calls to the latter. Furthermore, our method increases the degree of parallelism compared to other solutions that are based on a concatenation of sequential decoders, while maintaining similar logical resources (FPGA area) as a single BP decoder and minimizing latency.

The approach begins with a BP decoder with higher accuracy. In the event of a decoding failure, we activate two additional decoders: one that utilizes a scaling factor that provides a LER closer to the optimal BP, and another that introduces more diversity, following a tree structure. The goal is to tackle noisy scenarios by deliberately selecting non-optimal parameters to increase the number of decodable cases, similarly to the strategy proposed in \cite{shutty2024efficientnearoptimaldecodingsurface} for MWPM.

Once we activate the two decoders, we take the hard decision from the decoder that encountered a convergence failure and use it to adjust the a priori information for the subsequent decoder. 

We define the a priori information vector $\mathbf{y}$ as the log-likelihood ratio, $y_j = \log \left( \frac{\mathbb{P} (e_j = 0)}{\mathbb{P}(e_j = 1)} \right)$ with $j \in \{1, \dots, n\}$ and $\mathbf{e} \in \mathbb{F}_2^n$ denote the binary error vector. The error vector estimated by the decoder is defined as $\mathbf{e'} \in \mathbb{F}_2^n$, which is the binary vector based on the hard decision of the computed messages.

To further enhance diversity, we apply different modification factors, denoted as $\gamma_i$, where $i$ indicates the decoder number, like $\gamma_i \times \mathbf{e'} + (1-\gamma_i) \times\mathbf{y}$.

If neither of the two decoders converges, we engage two additional decoders in the more diverse branch. The first of these is a BP-based decoder, while the second is a BP combined with either LSD or OSD, but with a reduced number of iterations. Similar to the previous step, we modify the a priori information based on the hard decision from the second decoder, incorporating the $\gamma_i$ factors.

In summary, the procedure is as follows: 

\begin{enumerate}
    \item We first run the BP (sum product) decoder. 
    \item If it diverges, we run two parallel implementations of BP: one with accurate min-sum and another that is more diverse, utilizing different scaling factors for the exchanged messages, $\alpha$.
    \item These two decoders merge the a priori information with the hard decision, applying distinct $\gamma_i$ factors.
    \item If both of these decoders also diverge, we activate the final stage of decoders, using as input the a priori information modified by the hard decision from the previous decoder along with the $\gamma_i$ factor. One of these two decoders will include a post-processing stage like LSD or OSD.
\end{enumerate}

\begin{figure*}[!t]
\centering
\includegraphics[width=0.75\textwidth]{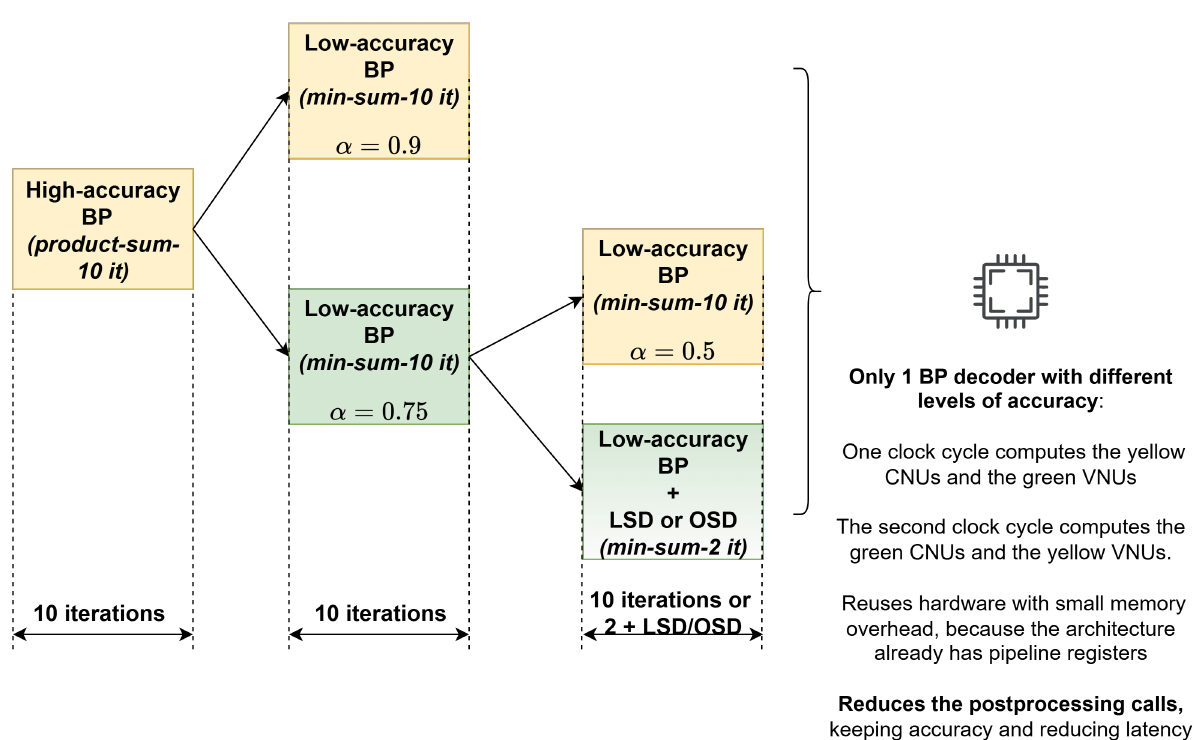}
\caption{Architecture for the diversity decoder based on BP implementations.}
\label{fig:DiversityCircuitResults}
\end{figure*}

\begin{figure}[h]
\centering
\includegraphics[width=0.5\textwidth]{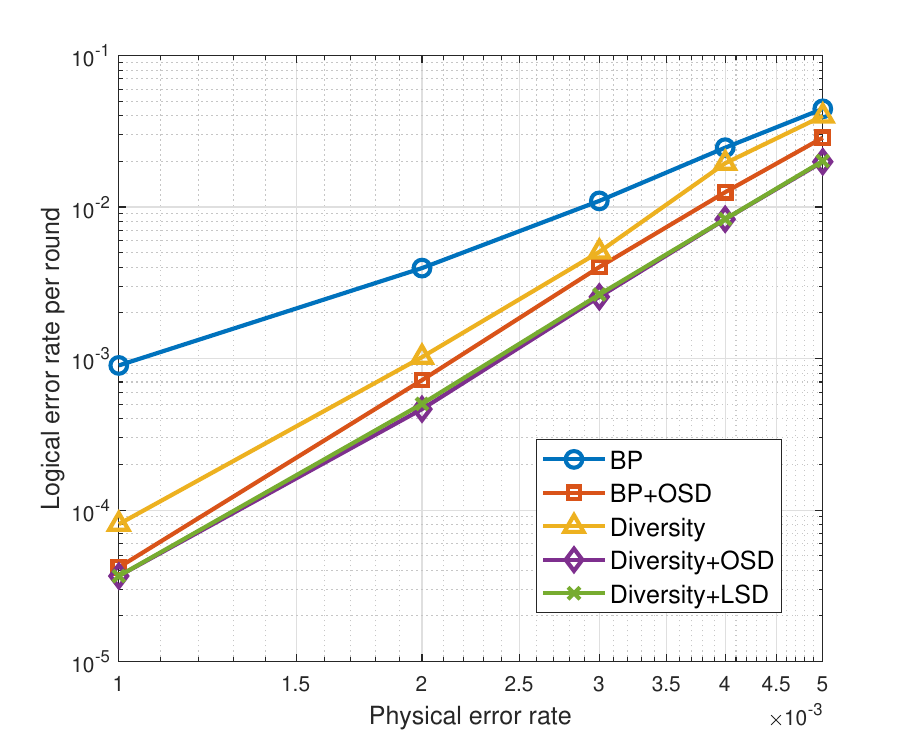}
\caption{LER for the bicycle bivariate code (72,12,6) under circuit level noise for the proposed diversity decoders and BP and BP+OSD.}
\label{fig:DiversityCircuitResults72}
\end{figure}

\begin{figure}[h]
\centering
\includegraphics[width=0.5\textwidth]{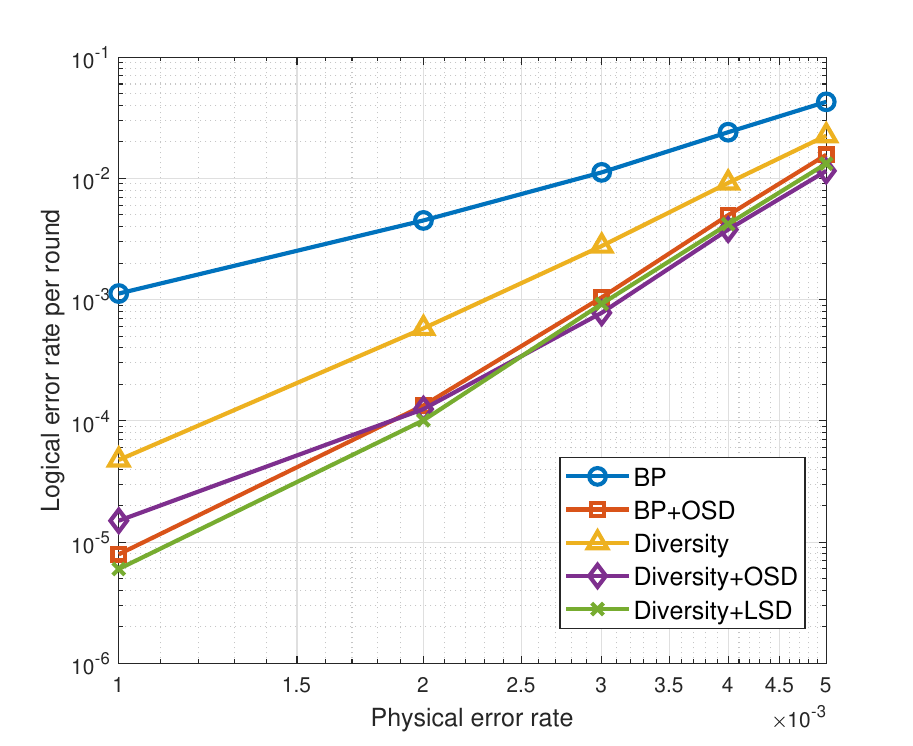}
\caption{LER for the bicycle bivariate code (108,8,10) under circuit level noise for the proposed diversity decoders and BP and BP+OSD. }
\label{fig:DiversityCircuitResults108}
\end{figure}

\begin{figure}[h]
\centering
\includegraphics[width=0.5\textwidth]{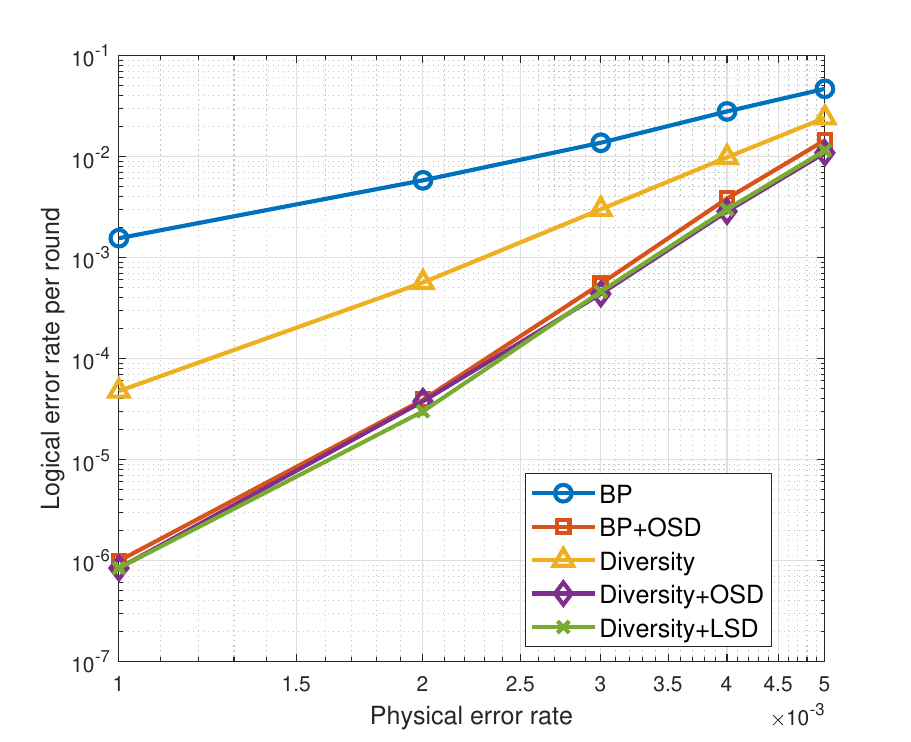}
\caption{LER for the bicycle bivariate code (144,12,12) under circuit level noise for the proposed diversity decoders and BP and BP+OSD.}
\label{fig:DiversityCircuitResults144}
\end{figure}

%\begin{figure}[h]
%\centering
%\includegraphics[width=0.5\textwidth]%{Figures/CircuitLevel.PNG}
%\caption{SOLO PARA EDICION: ACTUALIZAR CON LA CORRECTA.}
%\label{fig:DiversityCircuitResults}
%\end{figure}

We applied our proposed decoder to bivariate bicycle codes of different lengths: (72, 12, 6), (108, 8, 10), and (144, 12, 12) \cite{bravyi2024high}. The results were consistent across these simulations. For the diversity-based decoding process, we utilized a BP Sum-Product algorithm with 10 iterations as the base decoder. We included two Min-Sum decoders, each with scaling factors of $\alpha =$ 0.9 and 0.75, and $\gamma_0 = \gamma_1 =$ 0.75, also executing 10 iterations during the second stage. In the final stage, we employed one Min-Sum decoder with a scaling factor of $\alpha =$ 0.5, $\gamma_2 =$ 0.5  and 10 iterations, along with a combination of Min-Sum and LSD or OSD, with 2 iterations and $\gamma_3 =$ 0.5. \textcolor{black}{The values of the scaling factors and coefficients used in the diversity decoder are chosen to enable a hardware-friendly implementation rather than arbitrarily. These values allow the decoder to avoid hardware multipliers by implementing the products using simple bit shifts and additions. This design minimizes hardware resource usage and the latency while maintaining effective scaling.} This setup is summarized in Fig. \ref{fig:DiversityCircuitResults}. 

The worst-case latency for our approach is 22 iterations of BP plus the latency of LSD or OSD, \textcolor{black}{as 30 BP iterations of the parallel schedule is less restrictive.}

The baseline BP is executed for 100 iterations, while the benchmark BP+OSD is carried out following these 100 iterations of BP. All the simulations are performed under circuit-level noise as described in \cite{iolius2024almostlineartimedecodingalgorithm} \textcolor{black}{and for the LSD implementation, we refer to the authors' library \cite{Roffe_LDPC_Python_tools_2022}}.

In all cases analyzed, as illustrated in Figures \ref{fig:DiversityCircuitResults72}, \ref{fig:DiversityCircuitResults108}, and \ref{fig:DiversityCircuitResults144}, the LER of the proposed decoder is at least equivalent to that of BP+OSD. However, our approach requires significantly fewer calls to OSD and fewer BP iterations, as we analyze next. 

The results show that applying diversity decoding without LSD or OSD already improves the LER by an order of magnitude compared to BP for a physical error rate of 0.001 in all cases. When the post-processing is activated, this improvement is reflected in a reduction in the number of LSD or OSD activations, as summarized in Table \ref{table:DiversitySavings}. This technique results in greater savings with larger codes and demonstrates better performance with increasing physical error rates. 

Maintaining the post-processor after the diversity
warrants having at least the same performance. The reduced impact on average speedup in relation to decreasing physical error rates is attributed to the numerous cases in which BP converges within a small number of iterations. Contrarily, in higher-noise environments, the average latency is primarily influenced by LSD or OSD time.

When analyzing the worst-case scenario, the latency decreases in line with the reduction in the number of OSD calls. It is well established that, in worst-case situations, the post-processing time predominates rather than the BP time. However, it is important to highlight that our diversity proposal results in running 70 fewer iterations than BP+OSD, leading to savings in both time and power consumption without losing accuracy.

Moreover, the proposal shows a high confidence level, as when convergence is detected by the decoders, the occurrence of logical errors is negligible. In fact, the number of logical errors reported after declaring convergence decreases as the code length increases, as will be shown next.
For the 72-length code, convergence occurs 99.45\% of the time, and the cases that achieve convergence and produce a wrong codeword (logical failure) are just 0.001\% with the most accurate decoder, 0.0054\% with the second most accurate, and 0.0008\% with the least accurate BP decoders. Overall, the diversity decoder converges to an incorrect code in just 0.007\% of the cases in which the system declared a convergence when the physical error rate is 0.001. As the noise levels rise, with a physical error rate of 0.003, the solution converges 94.35\% of the time, resulting in a total of 0.58\% failures upon convergence.
The code with a length of 108 converges between 98.62\% and 87.57\% for physical error rates ranging from 0.001 to 0.003, respectively. In this case, convergence to a wrong codeword was found only twice out of $10^7$ simulations at a physical error rate of 0.003. A similar trend was observed with larger codes; specifically, the code of length 144 converged between 97.49\% and 78.93\% of the time, and after conducting over $10^7$ simulations, no convergence to incorrect codewords was found for physical error rates between 0.001 and 0.003.

\begin{table*}[t!]
\caption{Reduction in the number of LSD or OSD executions,  and average and worst-case speedup in software}
\centering
\begin{tabular}{|c|c|c|c|c|}
\hline
Physical error rate & 0.001 & 0.002& 0.003 & 0.005 \\ \hline
(72, 12, 6) & 91.04\%  & 74.14\% & 53.91\% & 47.77\% \\ \hline
(108, 8, 9) & 95.76\%  & 87.12\% & 75.31\% & 47.39\% \\ \hline
(144, 12, 12) & 96.93\%  & 90.27\%  & 78.08\% & 48.41\% \\ \hline
\hline
Average speedup & 1.3     & 1.5     & 1.6     & 1.8     \\ \hline
Worst-case speedup & 2.2     & 2.0     & 1.4    & 1.1     \\ \hline

\end{tabular}

\label{table:DiversitySavings}

\end{table*}

Compared to other existing decoders based on BP, such as SI and CA, our approach eliminates the need to calculate the reliabilities of checks in a graph, which can be complex in a detector error model. Additionally, we do not require an extra sorting process to arrange information based on reliability, significantly reducing latency. Due to the nature of BP, there is also no need to implement a consensus process among the decoders to enhance confidence like in \cite{shutty2024efficientnearoptimaldecodingsurface}.

When compared to the BP+BP+OSD method presented in \cite{iolius2024almostlineartimedecodingalgorithm}, our diversity decoder combined with OSD saves 100 iterations over the sparsified detector error model. The diversity proposal also achieves a larger performance gain compared to BP+BP without the need for two separate graphs, which can be important from a hardware standpoint. 

With our proposed diversity decoder, all arithmetic resources can be shared, requiring only the duplication of storage resources. In contrast, when using two graphs, it is not possible to share arithmetic and routing resources effectively. This means that with just one decoder, we can implement all three stages of the diversity decoder without sacrificing the degree of parallelism, as two decoders can operate simultaneously within the same architecture. 

\textcolor{black}{In comparison with recent competitive decoders, such as Relay-BP \cite{müller2025improvedbeliefpropagationsufficient}, our proposed diversity method offers a different trade-off. For the [[144,12,12]] QLDPC code, Relay-BP, which uses 301 relay legs, achieves a significant improvement in decoding performance, reducing the logical error rate by nearly an order of magnitude compared to BP+OSD. In contrast, our diversity approach, using only four BP decoders and a post-processor, results in a more modest improvement over BP+OSD, but it requires fewer decoders and still provides a 2.2x speedup in the worst case. While our solution does not improve BP+OSD as significantly as Relay-BP, it offers a more compact approach that reduces the number of decoding instances needed, making it a promising option for scenarios where minimizing computational resources is critical. Furthermore, combining our diversity approach with other methods such as Relay-BP, or SI could lead to a ``sweet spot'' that balances decoding performance and computational efficiency, providing a more flexible solution for different application needs and constraints.}

Although this diversity-based approach requires further exploration across more code families and configuration parameters, the current results suggest promising gains, either standalone or in combination with other methods. A fully normalized comparison in terms of hardware cost and resource usage is left for future work.

\section{Conclusion}
We have designed a hardware emulator implemented on an FPGA that enables us to explore, within days, the lower LER region necessary for implementing fault-tolerant quantum computation. This architecture is sufficiently flexible to run real QEC architectures while also monitoring and storing various parameters and error patterns. The platform is vendor-independent, and the results obtained have been instrumental in designing diversity-based decoding methods to enhance both convergence and accuracy. These diversity methods are based on analyzing and leveraging quantization noise, as well as employing various low-latency BP implementations. 

We analyze the results of both diversity methods under code-capacity and circuit-level noise, respectively, demonstrating significant improvements in speed and accuracy. These examples illustrate the value of examining the decoding problem from a hardware perspective, through hardware emulation, considering the specific characteristics of the architectures as part of the co-design process.

%Future work will focus on analyzing both diversity proposals in tandem, combining the diversity based on quantization noise with BP implementations to further reduce the number of LSD or OSD executions. Additionally, we will investigate the behavior of LSD and OSD after filtering a larger number of error patterns through diversity, with the goal of simplifying the algorithms to create more efficient and scalable implementations.

\textcolor{black}{Future work will focus on analyzing both diversity proposals in tandem, combining the diversity based on quantization noise with BP implementations to further reduce the number of LSD or OSD executions. Additionally, we will investigate how the diversity approach can be integrated with other ensemble-based methods, including Relay-BP, to reduce the number of BP decoders required while maintaining or improving performance. We also plan to explore the behavior of LSD and OSD after filtering a larger number of error patterns through diversity, with the goal of simplifying these algorithms and making them more efficient and scalable for future quantum computing applications.}

\section{Acknowledgment}

This work was supported by the QuantERA grant EQUIP (Spain MCIN/AEI/10.13039/501100011033, grant PCI2022-132922), funded by Agencia Estatal de Investigación, Ministerio de Ciencia e Innovaci\'{o}n, Gobierno de España and by the European Union ``NextGenerationEU/PRTR''. This research is part of the project PID2023-147059OB-I00 funded by MCIU/ AEI/ 10.13039/501100011033/ FEDER, UE. F. Garcia-Herrero's work on this project was partially funded by a grant from Google Quantum AI.

%Future work includes evaluation of the performance of the proposed decoding scheme under the circuit noise model for syndrome calculation, and combining the SB-LP post-processor with a space-time decoder that uses repeated noisy syndrome measurement, to asses possible performance-latency trade-offs in those computationally more demanding schemes.}

%This technique can potentially be employed in subsequent research for other belief propagation-based approaches which could be applied to other BP-based solutions in future work.

%Comparative analyses and simulations underscored its suitability for mitigating bit-flip and phase-flip errors. Our approach, using the SB-LP decoder as a post-processing step after the SB-MS decoder, showcased improved decoding performance and convergence efficiency. The detected shift in depolarizing probability and logical error rates highlights the algorithm's capacity to adapt to different error scenarios. 

 %\textcolor{black}{SUGGESTION (FRAN): Please put the titles of the BibTeX between double keys to keep the format of the tile.}

%\section*{Acknowledgement}
%This publication has emanated from research conducted with the financial support of Science Foundation Ireland under Grant Number \textcolor{red}{XXX}.
%This work is supported by the QuantERA grant EQUIP (Spain MCIN/AEI/10.13039/501100011033, grant PCI2022-132922), funded by Agencia Estatal de Investigación, Ministerio de Ciencia e Innovación, Gobierno de España and by the European Union “NextGenerationEU/PRTR”.
%\vspace{-0.3cm}

\bibliographystyle{IEEEtran}
\bibliography{{reference}}

\end{document}